\documentclass[aps, showkeys,prd]{revtex4}

\usepackage{amssymb}
\usepackage{color}
\usepackage{subfig}
\usepackage{float}

\usepackage{graphicx}
\usepackage{amsmath}
\usepackage{soul}
\usepackage[normalem]{ulem}

\usepackage{amsfonts}

\usepackage[english]{babel}
\usepackage{amsfonts}
\usepackage{amssymb}

\newcommand{\beq}{\begin{equation}}
\newcommand{\eeq}{\end{equation}}

\newcommand{\barr}{\begin{eqnarray}}
\newcommand{\earr}{\end{eqnarray}}

\newcommand{\gv}[1]{\ensuremath{\mbox{\boldmath$ #1 $}}}

\begin{document}

\title{Effective electromagnetic actions for Lorentz violating theories exhibiting the axial anomaly }
\author{Andr\'es G\'omez} \email{andresgz@ciencias.unam.mx}  
\affiliation{Facultad de Ciencias, Universidad Nacional Aut\'onoma de M\'exico, \\ 04510 M\'exico, Distrito Federal, M\'exico}
\author{A. Mart\'\i n-Ruiz}\email{alberto.martin@nucleares.unam.mx}
\author{ Luis F. Urrutia}
\email{urrutia@nucleares.unam.mx}
\affiliation{Instituto de Ciencias Nucleares, Universidad Nacional Aut\'onoma de M\'exico, \\ 04510 M\'exico, Distrito Federal, M\'exico}

\begin{abstract}
The CPT odd contribution to  the effective electromagnetic action deriving  from the vacuum polarization tensor in  a large class of fermionic systems exhibiting Lorentz invariance violation (LIV) is calculated using thermal field theory methods,  focusing upon  corrections depending on the chemical potential. The systems considered exhibit the axial anomaly and their effective actions are described by axion electrodynamics
whereby all the LIV parameters enter in the coupling $\Theta(x)$ 
to the unmodified  Pontryagin  density. A preliminary application to type-I tilted Weyl semimetals is briefly presented.
\end{abstract}

\keywords{Lorentz violation, Vacuum polarization, Thermal field theory, Axial anomaly}
\maketitle

\section{Introduction}

\label{INTRO}

The study of quantum corrections in the QED fermionic sector of the Standard Model Extension (SME)~\cite{Kostelecky0,Kostelecky1} due to the  electromagnetic interaction was sparkled in  Ref. \cite{Kostelecky1} which posed the problem of obtaining corrections to the   Chern-Simons interaction $\frac{1}{2} \epsilon ^{\mu\nu\alpha\beta} {\Delta k}_\mu A_\nu F_{\alpha\beta}$ as arising   from the one loop vacuum polarization tensor of fermions with the additional coupling  ${\bar \Psi} \gamma^5 \gamma^\mu b_\mu \Psi$.  Since then, a large number of authors have carried the  calculation of the 
CPT odd contribution to the vacuum polarization
obtaining a result that can be presented as $\Delta k^\mu= \zeta\, (e^2/\pi^2) \, b^\mu $, where $\zeta$ is always  a finite coefficient   adopting  different values, including $\zeta=0$, according to the regularization prescription chosen to deal with the superficially linear divergence of the vacuum polarization tensor $\Pi^{\mu \nu}$ \cite{CG,JK,PV,CHUNG1,CHUNG2,CHEN,BONNEAU,CHAICHIAN,PV1, BATTISTEL,ANDRIANOV,BALT1,BALT2,ALFARO}.  A survey of the principal approaches regarding  this issue can be found in Refs. \cite{PV1,ALFARO,JACKIW2,BALT3}.  The search of a physical criterion to select one specific value of $\zeta$ has been unsuccessful and the current understanding is that  $\zeta$ would be fixed either by  an experimental condition or by a more fundamental theory.  In order to briefly summarize  some of the previous results we assume that $b_\mu$ is timelike. The case $\zeta=0$ is obtained by demanding gauge invariance of the effective Lagrangian density \cite{CG}, and  has the drawback of eliminating  the Chern-Simons contribution from the outset since this term changes by a total derivative under gauge transformations.
Using the full  non-perturbative expression in $b_\mu$ for the fermion propagator and different regulators consistent with symmetric integration in four dimensions, the authors of Refs. \cite{JK,PV,CHUNG1,CHUNG2} find $|\zeta|={3}/{16}$. In an attempt to elucidate a further significance of the non-perturbative approach, which demands the use of  a unique regulator for all the contributions of $\Pi^{\mu\nu}$,  the  CPT even contribution to $\Pi^{\mu\nu}$  (second order in $b_\mu$) was calculated  in the $m=0$ case, finding that gauge invariance is violated \cite{BALT1}. The use of Pauli-Villars regularization was subsequently introduced in the non-perturbative calculation of the massive case, yielding gauge invariance to second order but two possibilities for the CPT odd contribution: either the expected value $\zeta=0$ or  $|\zeta|=3/8 $ \cite{BALT2}. Again, there is no unique way of fixing the $\zeta$ contribution by demanding gauge invariance of $\Pi^{\mu\nu}$.  An alternative regularization prescription based upon the maximal residual symmetry group of the vector $b_\mu$,  proposed in Ref. \cite{ANDRIANOV},  provides a 
definition on how to perform the loop integration  $ d^4 k $ 
 in $\Pi^{\mu\nu}$. For example, in the timelike case, the maximal residual symmetry group is $SO(3)$ which calls for  a spherically symmetric integration in the tri-momentum ${\mathbf k}$ instead of on the four-momentum $k^\mu$. This prescription yields $|\zeta|=1/4$.

In this work we extend previous calculations of effective electromagnetic action induced by radiative correction in the SME by including additional terms of the fermionic sector in the minimal QED extension of the SME \cite{TABLE}. Our starting point is the  action
\begin{align}
S = \int d ^{4} x \, \bar{\Psi} \left( \Gamma ^{\mu} i \partial _{\mu} - M
- e \Gamma ^{\mu} A _{\mu} \right) \Psi,  \label{ACTION1}
\end{align}
which is coupled to the electromagnetic field $A^\mu= (A^0, A^i)= (A^0, {\gv A})$, with  the notation   ${\gv A}=(A_x,\, A_y, \, A_z)$ in terms of the Cartesian components.  The  metric is $\eta_{\mu\nu}= \mbox{diag}(+,-,-,-)$ and we use  $\hbar = c = 1$ units henceforth. We set    $m=m_5=H_{\mu\nu}=e_\lambda=f_\lambda=g_{\lambda \kappa \nu}=0 $,  in  the notation of Table XVI of Ref. \cite{TABLE}. Thus we  restrict ourselves to 
\begin{equation}
\Gamma ^{\mu }= c^{\mu }{}_{\nu }\gamma ^{\nu }+d^{\mu
}{}_{\nu }\gamma ^{5} \gamma ^{\nu },\qquad M=a_{\mu }\gamma ^{\mu }+b_{\mu
}\gamma ^{5} \gamma ^{\mu },
\label{GAMMAMU}
\end{equation}%
with $\gamma ^{\mu }$ being  the standard gamma matrices. The term $c^\mu{}_\nu$ already includes the  $\delta^\mu_\nu$ contribution corresponding to the free Dirac action.  The terms ${\bar \Psi}\, \Gamma^\mu i \partial_\mu \Psi$  are CPT even, while those in ${\bar \Psi} M \Psi$ are CPT odd. Still, since $\gamma^5$ is PT odd,  each of them separately contains a mixture of PT even and odd terms.

Let us emphasize that the methods and techniques employed in the  calculation of these extended effective actions in high energy physics, besides being valuable in their own, can be of relevance in some areas of condensed matter physics.  Indeed,  the identification of fermionic quasiparticles of Dirac and/or Weyl type in the linearized approximation of Hamiltonians in topological phases of matter provides the opportunity of studying them under the perspective of the SME.  Such approach has been particularly fruitful in the case of Weyl semimetals (WSMs) whose electronic Hamiltonians naturally include some of the LIV terms considered in the fermion sector of the SME.  Nevertheless, in this case, the LIV parameters need not be highly suppressed, since they are determined by the electronic structure of the material and are subjected to experimental determination                
\cite{GRUSHIN,Burkov,Goswami,TI,REVLANDS, Kostelecky3}.  Our choice of the LIV parameters in Eq. (\ref{GAMMAMU}) are those relevant for the description of a general WSM. The simplest example arises in the Hamiltonian of a Weyl semimetal with no tilting and with an isotropic Fermi velocity,  which can be embedded in
the fermionic action
\begin{align}
S = \int d ^{4} x \, {\bar \Psi}(x) \left( i \gamma ^{\mu} \partial _{\mu} - {{\tilde b}_{\mu}  \gamma ^{5}  \gamma ^{\mu} } \right) \Psi(x), \label{1}
\end{align}
where {${\tilde b}^{\mu} = ( {\tilde b}^{0} , {\gv{{\tilde b}}} )$}.  Starting from  the  action (\ref{1}) coupled to an electromagnetic field,  the chiral rotation method of  Ref.  \cite{Burkov},  which provides an alternative quite different from the standard vacuum polarization calculation,  yields
\begin{align}
S _{{\rm{eff}}} = \frac{e ^{2}}{32 \pi ^{2}} \int d ^{4}x \,  \Theta(x) \,  \epsilon^{\alpha \beta \mu \nu} F _{\alpha \beta} F_{\mu \nu}, \quad \Theta(x) = 2 {\tilde b}_{\mu} x ^{\mu},  \label{02}
\end{align}
which determines $|\zeta|=1/4$.  Another outstanding example relevant in condensed matter is the case of the superfluid $^3$He-A, where the value $\zeta=1/2$  is reported \cite{VOLOVIK}. 

The action (\ref{02}) captures the electromagnetic response of WSMs. While the axion coupling $\Theta(x)$ arises from the nontrivial topology of the band structure of the material, the appearance of the abelian Pontryagin density (APD) $\epsilon^{\alpha \beta\mu\nu} F_{\alpha \beta} F_{\mu\nu}$ is related to the axial anomaly
\begin{align}
\partial ^{\mu} J _{\mu} ^{5} = - \frac{e ^{2}}{16 \pi ^{2}} \, \epsilon ^{\alpha \beta \mu \nu} F_{\alpha \beta} F_{\mu\nu} . 
\end{align}
From Eq.  (\ref{02}) we obtain the effective sources \cite{FRANZ}
\beq
\rho=\frac{\delta S_{{\rm eff}}}{\delta A_0}=\frac{e^2}{2 \pi^2}{\gv {\tilde b}}\cdot{\gv B},\qquad {\gv J}=\frac{\delta S_{{\rm eff}}}{\delta {\gv A}}=\frac{e^2}{2 \pi^2}({\gv {\tilde b}}\times{\gv E}-{\tilde b}_0 {\gv B}),\label{EFS}
\eeq 
where ${\gv E}$ and ${\gv B}$ are the electromagnetic fields. In our index notation the above current is 
$J_i=\delta S_{\rm eff}/\delta A^i$. In particular, for ${\tilde b}_0=0$,  this yields 
$J_x=\sigma_{xy} E_y$ with the transverse anomalous Hall conductivity $\sigma_{xy}=-\frac{e^2}{2 \pi^2}{\tilde b}_z$, a distinguishing feature of Weyl semimetals \cite{BURKOVBALENTS}.   As it will become clear in Section \ref{MUINDEP}, the knowledge of this quantity allows to unambiguously fix the otherwise arbitrary coefficient $\zeta$ in this case. This additional information  is a general consequence of the underlying  microscopic theory describing the material, as opposed to the lack of a more  fundamental theory containing the SME.

 Recently, the appearance of the APD in effective actions  arising from (\ref{ACTION1}) has promoted the use of anomaly calculations to obtain them. For example, in the path integral approach,  $S_{\rm eff}$ was obtained  by introducing the electromagnetic coupling in  Eq.~(\ref{1}) and subsequently eliminating the fermionic term proportional to ${\tilde b}_\mu$ through a chiral rotation.  Nevertheless, this produces an  electromagnetic contribution to the action arising from the nonzero Jacobian of the chiral rotation which is proportional to the Pontryagin density~\cite{Burkov}. Following this idea, the Fujikawa prescription to obtain the chiral anomalies \cite{Fujikawa,Bertlement} has also been used to calculate the effective electromagnetic action of different materials  in Refs. \cite{Goswami,TI,REVLANDS} as well as to give an alternative explanation of the indeterminacy of the coefficient $\zeta$ related to some freedom in the definition of the fermionic axial vector current \cite{CHUNG3}.

Nevertheless, as pointed out in Ref. \cite{LUAGA}, the anomaly does not directly yield the effective action. Also,  the method of eliminating the modifying fermionic contributions via a chiral rotation cannot be  easily extended to deal with the more complicated configurations envisaged  in the action (\ref{ACTION1}). 

The reasons indicated above suggest the convenience of applying and extending the    quantum field theory methods developed in high energy physics  to obtain  the required effective electromagnetic actions corresponding to fermionic systems described by the action (\ref{ACTION1}) which are of interest in condensed matter physics.   In particular this procedure  should clarify how  the LIV corrections enter in the effective action, while the chiral anomaly remains insensitive to them \cite{LUAGA, FIDEL,SALVIO,Scarpelli}.  Motivated by the inclusion of  temperature effects in  the odd contribution of the vacuum polarization tensor arising from  the $b_\mu$ coupling in the action (\ref{1}), reported  in Refs. \cite{EBERT, TEMP1, TEMP2, TEMP3, TEMP4, TEMP5},  we provide the first steps to incorporate thermal field theory in the case of the more general LIV couplings defined  in Eqs. (\ref{GAMMAMU}).  In this way, we incorporate non-zero chemical potential effects,  but still remain in the zero temperature limit. We follow the conventions of Ref.~\cite{PESKIN}.

\section{ The effective action}

Integrating the fermions in Eq.  (\ref{ACTION1}) defines  the effective action
\begin{align}
\exp( i S _{\rm eff} ) = \det \Big[ {\bar \Psi} (x) \, \Big( \Gamma ^{\mu} i \partial _{\mu} - M - e \Gamma ^{\mu} A _{\mu} \Big) \, \Psi (x) \Big],  \label{9.0}
\end{align}
which we write as
\begin{align}
S _{\rm eff} ^{(2)} (A) = \frac{1}{2} \int \frac{d ^{4} p}{ \left( 2\pi \right) ^{4}} \, A _{\mu} (-p) \, \Pi ^{\mu \nu }(p) \, A_{\nu} (p) ,  \label{9}
\end{align}
to second order in the electromagnetic potential. This  introduces  the vacuum polarization tensor $\Pi^{\mu\nu}(p)$
\begin{align}
i \Pi ^{\mu \nu} (p) = e ^{2}  \int \frac{d^{4}k}{\left( 2\pi \right) ^{4}} \, \mbox{tr} \left[ S(k-p)\Gamma ^{\mu }S(k)\Gamma ^{\nu } \right]   , \label{PIMUNU}
\end{align}
where $S(k) = i / ( \Gamma ^{\mu} k _{\mu} - M )$ is the exact fermion propagator in momentum space including LIV modifications. Since $S _{\rm eff} ^{(2)}$ is real we must have $\Pi^{*}_{\mu\nu}(p)=\Pi_{\nu\mu}(p)$. As we will show, the new effective action (\ref{9}) will keep the form of Eq. (\ref{02}) with all the modifications entering through a new vector ${\cal B}_\lambda$ to be determined.
The link between both expressions is accomplished with the identifications
\begin{align}
\Pi ^{\mu \nu} (p) = -i \frac{e ^{2}}{2 \pi ^{2}} \, {\cal B} _{\lambda} p _{\kappa} \epsilon ^{\mu \nu \lambda \kappa} , \quad \Theta (x) = 2 {\cal B} _{\lambda}  x ^{\lambda} . \label{BLAMBDA}
\end{align}
In the following we calculate the CPT odd contribution of the effective action (\ref{9}). In the massless case ($m=0$), $\Gamma ^{\mu} k _{\mu} - M $ is linear in $\gamma ^{\mu}$ and $\gamma ^{5} \gamma ^{\mu}$. The appearance of the matrix $\gamma ^{5}$ suggests the convenience of using  left and right chiral projectors in order to evaluate $\gamma^5=\pm 1$ \cite{BALT1,SALVIO, TEMP1}. Therefore, we can work in the chiral basis, where the decomposition of the Dirac spinor into the right and left Weyl spinors is manifest.  These latter are eigenspinors of $\gamma ^{5}$ with the eigenvalues $\pm 1$.  We now define the projection operators
\begin{align}
P _{R} = \frac{ 1 + \gamma ^{5} }{2} , \qquad P _{L} = \frac{ 1  -  \gamma ^{5} }{2}, \qquad \gamma _{5} ^{2} = 1  ,  \label{11}
\end{align}
which project onto right- and left-handed spinors, respectively. Note that $\gamma ^{\mu} P _{L} = P _{R} \gamma ^{\mu}$. The projectors (\ref{11}) allow us to define the matrices $\Gamma _{R} ^{\mu}$ such that
\begin{align}
\Gamma ^{\mu} P _{R} = \left( c ^{\mu} {}_{\nu} - d ^{\mu} {}_{\nu} \right) \gamma ^{\nu} P _{R} \equiv \Gamma _{R} ^{\mu} P _{R} , \label{12}
\end{align}
which identifies $ \Gamma ^{\mu} _{R} = r ^{\mu} {}_{\nu} \gamma ^{\nu}$ with $r ^{\mu} {}_{\nu} = c ^{\mu} {}_{\nu} - d ^{\mu} {}_{\nu}$. In an analogous way we define the left-handed part of the matrices $\Gamma ^{\mu}$ as
\begin{align}
\Gamma ^{\mu} P _{L} = \Gamma _{L} ^{\mu} P _{L} , \qquad \Gamma _{L} ^{\mu} = l ^{\mu} {}_{\nu} \gamma ^{\nu} ,\qquad l ^{\mu} {}_{\nu} = c ^{\mu} {}_{\nu} + d ^{\mu} {}_{\nu} \, .  \label{13}
\end{align}
Following the same idea, we can also split the fermion propagator $S (k)$ into its right- and left-handed parts, i.e.
\begin{align}
\frac{i}{\Gamma ^{\mu} k _{\mu} - M } \gamma ^{\alpha} P _{R} = P _{R} S _{R} (k) \gamma ^{\alpha}, \quad  S _{R} (k) = \frac{i}{\big( k _{\mu} r ^{\mu}{}_{\nu} - a _{\nu} + b _{\nu} \big) \gamma ^{\nu}} , \notag \\  \frac{i}{\Gamma ^{\mu} k _{\mu} - M } \gamma ^{\alpha} P _{L} = P _{L} S _{L} (k) \gamma ^{\alpha} , \quad  S _{L} (k) = \frac{i}{\big( k _{\mu} l ^{\mu}{}_{\nu} - a _{\nu} - b _{\nu} \big) \gamma ^{\nu}}.  \label{14}
\end{align}
Note that the propagators $S_{L}$ and $S_{R}$, having the generic form $i/(Z_\nu \gamma^\nu)$, can be readily rationalized as $i (Z_\nu \gamma^\nu)/ Z^2 $.

To proceed forward with the calculation we now split the combination $T ^{\mu \nu} (k,p) = S(k-p) \Gamma ^{\mu} S(k) \Gamma ^{\nu}$ under the trace in $\Pi ^{\mu \nu }(p)$ into its left- and right-handed parts, i.e. 
\begin{align}
T _{L(R)} ^{\mu \nu } (k,p) &= S(k-p) \Gamma ^{\mu} S(k) \Gamma ^{\nu} P _{L(R)} , \label{18}
\end{align}
which implies that the vacuum polarization can be written as the sum $\Pi ^{\mu \nu} (p) = \Pi _{L} ^{\mu \nu} (p) + \Pi _{R} ^{\mu \nu} (p)$, where 
\begin{align}
i \Pi _{L(R)} ^{\mu \nu} (p) &= e ^{2} \int \frac{d ^{4} k}{\left( 2 \pi \right) ^{4}} \, \mbox{tr} \left[ T _{L(R)} ^{\mu \nu} (k,p) \right]  \label{17} 
\end{align}
is the vacuum polarization for a left-(right-)handed massless fermion. We now concentrate in the calculation of $\Pi _{L(R)} ^{\mu \nu} (p)$. Using Eqs. (\ref{12})-(\ref{14}) and the cyclic property of the trace, the left-handed part $\Pi _{L} ^{\mu\nu}$ can be written as
\begin{align}
 i \Pi _{L} ^{\mu \nu} (p) &= e ^{2} l ^{\mu} {}_{\beta} l ^{\nu} {}_{\alpha} \int \frac{d^{4}k}{\left( 2\pi \right) ^{4}} \, \mbox{tr} \left[ S _{L} (k-p) \gamma ^{\beta}  S _{L} (k) \gamma ^{\alpha} P _{L} \right] . \label{19}
\end{align} 
A similar procedure yields the right-handed part $\Pi ^{\mu \nu} _{R} (p)$ by means of the replacements $L \to R$, $l ^{\mu} {}_{\nu} \to r ^{\mu} {}_{\nu}$, $b _{\mu} \to - b _{\mu}$, $P _{L} \to P _{R}$ in Eq. (\ref{19}). In the following we restrict ourselves to the axial contributions  $\Pi ^{\mu \nu} _{A,L}$ and $\Pi ^{\mu \nu} _{A,R}$ of the left- and right-handed terms, respectively, which are obtained by isolating the terms  $P _{R} \to + \gamma ^{5} /2$ and $P _{L} \to - \gamma ^{5} /2$ in the corresponding expressions for the vacuum polarization. Both expressions can be summarized in  the general  form
\begin{align}
i \Pi _{A, \, \chi } ^{\mu \nu} (p) &= \frac{{\chi}}{2} e^{2} m^{\mu}{}_{\beta} m ^{\nu}{}_{\alpha} \int \frac{d ^{4} k}{\left( 2 \pi \right) ^{4}} \mbox{tr} \left[ S _{{\chi}} (k-p) \gamma ^{\beta} S _{{\chi}} (k) \gamma ^{\alpha} \gamma ^{5} \right] ,   \label{21}
\end{align}
with the following assignments 
\begin{eqnarray}
R &:&  {\chi = + 1} \qquad m ^{\mu} {}_{\nu} = r ^{\mu}{}_{\nu} , \qquad  C _{ \rho} = a _{\rho} - b _{\rho} , \nonumber \\
L &:& {\chi = - 1} \qquad m ^{\mu}{}_{\nu} = l ^{\mu}{}_{\nu} , \qquad C _{ \rho} = a _{\rho} + b _{\rho} . \label{22}
\end{eqnarray}
Clearly, the full axial contribution to the vacuum polarization is the sum of the $L$ and $R$ parts, i.e. $ \Pi _{A} ^{\mu \nu} = \Pi _{A, L} ^{\mu \nu} + \Pi _{A, R} ^{\mu \nu}$. For the sake of simplicity, we do not introduce an additional subindex $\chi$ either in the matrix $m^\alpha{}_\beta$ or in the vector $C _{\rho}$, which are to be restored at the end of the calculations according to Eq. (\ref{22}). The calculation in Eq. (\ref{21}) proceeds as follows: to simplify the notation we introduce the primed vectors $q _{\nu} ^{\prime} = q _{\mu} m ^{\mu}{}_{\nu}$ maintaining the original measure $d ^{4} k$. Also we  rationalize the denominators in the propagators and  take the trace $\mbox{tr} (\gamma ^{\rho} \gamma ^{\beta} \gamma ^{\sigma}\gamma ^{\alpha} \gamma ^{5}) = - 4 i \epsilon ^{\rho \beta \sigma \alpha}$, with $\epsilon ^{0123} = + 1$. We  use the antisymmetry of the Levi-Civita symbol to eliminate the term $\epsilon ^{\alpha \beta \rho \sigma} (k ^{\prime} - C ) _{\rho} (k ^{\prime} - C)_{\sigma}$ in the numerator. Since we are interested in the contribution to the effective action including the product of two electromagnetic tensors without additional derivatives, we calculate the integral  Eq. (\ref{21}) only to first order in $p _{\alpha}$. Finally, and assuming that $m ^{\mu}{}_{\nu}$ is invertible, the identity $m ^{\mu}{}_{\beta} \, m ^{\nu}{}_{\alpha}\,  m ^{\kappa}{}_{\sigma} \epsilon ^{\beta \alpha \rho\sigma} = ( \det m) \,  (m ^{-1})^{\rho}{}_{\lambda} \epsilon ^{\mu \nu \lambda \kappa}$ yields the further simplification
\begin{align}
\Pi _{A, \, \chi }^{\mu \nu }(p) = - 2 {\chi} e ^{2} (\det m)(m^{-1}) ^{\rho} {}_{\lambda} \epsilon ^{\mu \nu \lambda \kappa} p _{\kappa} \, I _{\rho} (C),  \label{23}
\end{align}
where we have defined the integral
\begin{align}
I _{\rho} (C) = \int  \frac{d ^{4} k }{\left( 2 \pi \right) ^{4}} \, g _{\rho} (k _{0} , {\gv k}) \, , \quad  g _{\rho} (k _{0} , {\gv k}) = \frac{ \left( k' - C \right) _{\rho} }{ \left[ \left( k' - C \right) ^{2}\right] ^{2} }, \label{24}
\end{align}
and  we recall that ${k'}_\mu={k}_\alpha m^{\alpha}{}_\mu$. 
Previous to regularization, the above expression is our final result for the vacuum polarization tensor in Minkowski spacetime, which can be evaluated for the left- and right-handed fermions according to Eq. (\ref{23}) with the assignments in Eq. (\ref{22}). The result (\ref{23}) holds for arbitrary LIV terms $c^\mu{}_\nu, \, d^\mu{}_\nu, \, a_\mu$  and $b_\mu$ as long as these produce invertible matrices $l^{\mu}{}_\nu$ and $r^{\mu}{}_\nu$.

\section{The regularization}

As a first step in the inclusion of  thermal field theory methods in the description of the radiative corrections to the  fermionic action (\ref{ACTION1}) under the new conditions (\ref{1}), here we shall consider the case of a non-zero chemical potential $\mu$ but still remain in the zero-temperature limit.  The inclusion of both will be published elsewhere.  To this end we adopt the finite temperature approach in the imaginary time  formulation \cite{KAPUSTA} , which we recover after the substitution \cite{BERNARD,DITTRICH,DAS} 
\begin{align}
\int\frac{d ^{4} k}{(2 \pi ) ^{4} } \ \  \rightarrow \ \  +i T \sum_ {n = - \infty} ^{\infty} \int \frac{d ^{3} {\gv k} }{(2 \pi ) ^{3}} , \label{25} 
\end{align}
with $k_{0}\to k_{0}=i\omega_{n}+\Lambda$. 

\begin{figure}[H]
\centering
\includegraphics[scale=0.52]{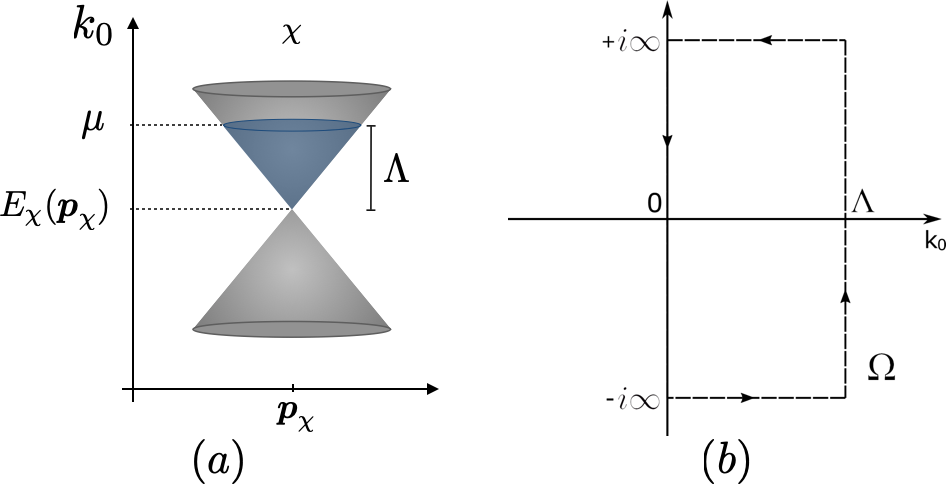}
\caption{(a) The effective chemical potential. (b) The contour of integration in the $k _{0}$-plane. }
\label{FIG1}
\end{figure}

In order to have a specific system which determines the choice of the finite but undetermined parameters, together with a better physical understanding of the subsequent steps, we introduce  the Hamiltonian
\begin{align}
H _{\chi} (\gv{p}) = {\gv v}_\chi \cdot ({\gv p}   +  \chi {\gv {\tilde b}})  -   \chi {\tilde b}_{0}  + \chi v _{F} {\gv \sigma} \cdot ({\gv p}  + \chi {\gv {\tilde b}}) , \label{03}
\end{align}
borrowed from condensed matter,  which describes a Weyl semimetal with two linear 3D band crossings of chirality $\chi = \pm 1$ close to {$\pm {\gv {\tilde b}}$} in momentum space and {$\pm{\tilde b} _{0}$} in energy. 
Here $v_{F}$ is the isotropic Fermi velocity at each band crossing, ${\gv \sigma} = (\sigma _{x} , \sigma _{y} , \sigma _{z})$ is the triplet of spin-$1/2$ Pauli matrices,  ${\gv v} _{\chi}$ is the tilting parameter and ${\gv p}$ is the momentum. In the Weyl basis for the gamma matrices
\begin{equation}
\gamma ^{0}=%
\begin{pmatrix}
0 & \sigma _{0} \\ 
\sigma _{0} & 0%
\end{pmatrix}%
,\hspace{0.5cm}\gamma ^{i}=%
\begin{pmatrix}
0 & \sigma ^{i} \\ 
-\sigma^{i} & 0%
\end{pmatrix}%
,\hspace{0.5cm}\gamma ^{5}=%
\begin{pmatrix}
- \sigma _{0} & 0 \\ 
0 & \sigma _{0} %
\end{pmatrix},%
\end{equation}
both $2 \times 2$ Hamiltonians in Eq. (\ref{03}) can be embedded in the action (\ref{ACTION1}) with the choices $c^0{}_0=1, c ^{0}{}_{i} = d ^{0} {}_{i} = d^0{}_0= 0$, i.e.  $\Gamma ^{0} = \gamma ^{0}$ together with  $\Gamma ^{i}$ and $M$ being determined by the parameters in (\ref{03}).  For simplicity, the anisotropy in the Fermi velocities is not included in (\ref{03}) but enters naturally in our final result through  the coefficients $c^i{}_j$ and $d^i{}_j$. 
The Hamiltonian (\ref{03}) 
is realized in terms of chiral fermions whose dispersion relation behaves linearly,  with band crossing points localized both in momentum and energy,  as schematically indicated in the Fig. \ref{FIG1}(a).  As usual,  the position of the chemical potential determines most of the transport properties of a given material.  In a metal for instance,  it has to be measured from the gap closing, since it represents the filling of either the conduction or the valence bands. This suggests that in WSMs, the transport properties will depend on the band filling,  i.e.  on the chemical potential as measured from the node of the cones.  Therefore, in order to apply the finite temperature approach (\ref{25}) to WSMs, the parameter $\Lambda$ in Eq. (\ref{25}) has to be understood as the chemical potential measured from the band-crossing points, as shown in Fig. \ref{FIG1}(a).  To be precise,
$\Lambda=\mu-E_{\chi}({\gv{p}}_{\chi})$, where ${\gv{p}}_{\chi}$ is the
location in momentum of the node with chirality $\chi$, and
$E_{\chi}({\gv{p}}_{\chi})$ the corresponding energy. Both
${\gv{p}}_{\chi}$ and $E_{\chi}({\gv{p}}_{\chi})$ will be determined later
for the problem at hand. In the simplified model (3), it is clear that
${\gv{p}}_{\chi}=\chi {\gv {\tilde b}}$ and $E_{\chi}({\gv{p}}_{\chi})= -   \chi {\tilde b}_{0}$.

The sum in Eq. (\ref{25}) is over the Matsubara frequencies $\omega _{n} = (2n+1) \pi T$ required to produce anti-periodic boundary conditions for the fermions \cite{KAPUSTA}. Next we focus in Eq. (\ref{24}) and we make use of the additional relation \cite{KAPUSTA} 
\begin{align}
 \lim _{T \to 0} T \sum _{n = - \infty} ^{\infty} g _{\rho} ( k _{0} = i \omega _{n} + {\Lambda} , {\gv k} ) & = \frac{1}{2 \pi i}\int _{- i \infty} ^{ i \infty} d k _{0} \, g _{\rho} ( k _{0} , {\gv k} ) \, + \notag \\ & \phantom{=} + \frac{1}{2 \pi i} \oint _{\Omega} d k _{0} \, g _{\rho} ( k _{0} , {\gv k} ), \label{Matsubara_fermion}
\end{align}
where the contour $\Omega$ is shown in the Fig. \ref{FIG1}(b).

\subsection{The $\mu$-independent contribution}
\label{MUINDEP}

The first term in the right hand side of Eq. (\ref{Matsubara_fermion}) reduces to  the standard  zero-temperature, zero-chemical potential contribution, which has been previously discussed  in the literature  as extensively reported in section \ref{INTRO}. Going back to Eq. (\ref{23}), this corresponds to the direct evaluation of the integral $I_\rho(C)$
after the change of integration variables $k ^{\prime} _{\mu} = k _{\nu} m ^{\nu}{}_{\mu}$. Since the only vector at our disposal is $C_\mu$ we have 
\beq
I^{(1)}_\rho(C)= \frac{i}{(\det m)}\,  {\bar N}  C_\rho, 
\quad {\bar N}=\frac{1}{C^2}\left[ \int \frac{d ^{4} k ^{\prime}}{\left( 2 \pi \right) ^{4}} \; \frac{ \left( k ^{\prime} - C \right) \cdot C}{\left[\left( k ^{\prime} - C \right) ^{2} \right] ^{2} } \right]_E , \label{26}
\eeq
where the integral inside the square brackets is in  Euclidean space and the factor $+i$ comes from the Wick rotation.
The factor ${\bar N}$ is regularization dependent and could only be a function of the magnitude of the four-vector $C _{\mu}$. However, a change of scale $C _{\sigma} \to \lambda C _{\sigma}$ followed by an additional change of variables  $k ^{\prime \prime} _{\mu} = \lambda k ^{\prime} _{\mu}$ shows that ${\bar N}$ is just a numerical factor, independent of $C _{\mu}$. Therefore, ${\bar N}$ is the same for both left- and right-handed fermions, i.e. ${\bar N} _{L} = {\bar N} _{R} = {\bar N}$. In this way, the total contribution to the vacuum polarization in this case is summarized in the four-vector
\begin{align}
{\cal B}^{(1)} _{\lambda} = -4 \pi ^{2} {\bar N} \, \Big[ C _{L  \rho} \, (l^{-1}) ^{\rho} {}_{\lambda} - C_{R \rho} \, (r ^{-1}) ^{\rho}{}_{\lambda} \Big], \label{29a} 
\end{align}
according to Eq. (\ref{BLAMBDA}). As shown previously in the literature the factor ${\bar N}$ is finite but undetermined. In the case of WSMs, its dependence  upon the regularization procedure has been studied in Ref. \cite{Goswami} and the final choice is  made  by selecting the anomalous Hall conductivity $\sigma_{xy}=-e^2 {\tilde b}_z/(2 \pi^2$) as the physical quantity to be reproduced in the zero-tilting limit of the Hamiltonian (\ref{03}) \cite{Burkov,Goswami,BURKOVBALENTS}.  To this end it is necessary to
take a cut-off in the direction of the spatial component of $C$,  with the result  $ {\bar N}=-1/(8\pi^2)$.  The suitability of this choice will be demonstrated later by comparing our field-theoretic results with those obtained from a semiclassical Boltzmann approach, which has proven to be successful in the study the transport properties of WSMs. 

\subsection{The $\mu$-dependent contribution}
Next we consider  the second term in the right-hand side of Eq. (\ref{Matsubara_fermion}) and calculate
\begin{align}
I ^{(2)} _{\rho} (C) = \int \frac{d ^{3} {\gv k}}{(2 \pi ) ^{3}} \, \frac{1}{2 \pi i} \, \oint _{\Omega} d k _{0} \,  g _{\rho} ( k _{0} , {\gv k})  .  \label{29}
\end{align}
Let us start by finding the double poles of the integrand in the $k _{0}$-plane. They are located at the two points $k ^{\prime} _{0 \pm} = C _{0} \pm \vert {\gv k} ^{\prime} - {\gv C} \vert $, where we recall the relations $k ^{\prime} _{0} = k _{\mu} m ^{\mu}{}_{0}$ and $k ^{\prime} _{i} = k _{0} m ^{0}{}_{i} + k _{j} m ^{j}{}_{i}$. In the general case the solution for $k_0$ involves a rather cumbersome second order equation which will not be very illuminating for our purposes. In order to illustrate the full procedure we  restrict ourselves to the simpler situation where $m ^{0}{}_{\nu} = \delta ^{0}_{\nu}$. Under  this assumption the poles are located at
\begin{align}
k _{0 \xi} = - k _{j} m ^{j}{}_{0} + C _{0} + \xi \,  \vert {\gv k} ^{\prime} - {\gv C}  \vert , \quad \xi = \pm 1 ,  \quad k ^{\prime} _{j} = k _{i} \, m ^{i}{}_{j} ,
\end{align}
which corresponds to the dispersion relation of the model, wherefrom we can determine the position  in momentum and energy of the Dirac/Weyl node. Clearly, the band touching point (i.e. the node) occurs at ${\gv k}^{\prime}={\gv C} $, and the corresponding energy is $E_{0}\equiv k_{0\xi}({\gv k} ^{\prime}={\gv C})$, which can be explicitly expressed as
\begin{equation}
E_{0} = - \mathcal{V} ^{j} C _{j} + C _{0}  = \gv{\mathcal{V}}\cdot {\gv C} + C _{0}, \quad \mathcal{V} ^{j} = (m ^{-1}) ^{j}{}_{i} \, m ^{i}{}_{0}. \label{Zero_Energy}
\end{equation}
Applying the residue theorem we obtain
\begin{align}
 \frac{1}{2\pi i} \oint _{\Omega} d k _{0} \, g _{\rho} = \sum _{  \xi = \pm 1 } \mathrm{Res}( g _{\rho}, k _{0} = k _{0 \xi}) H (k _{0 \xi}) H ( \Lambda - k _{0 \xi}), 
\end{align}
where the Heaviside functions {$H$} guarantee that the poles $k _{0 \xi}$ fall inside the contour $\Omega$ of Fig. \ref{FIG1}, i.e.  $0 < k _{0 \xi} < \Lambda$, where $\Lambda=\mu-E_{0}$, with $E_{0}$ given by Eq. (\ref{Zero_Energy}). The residues are 
\begin{equation}
\mathrm{Res}( g _{\rho}, k _{0} = k _{0 \xi}) = - \xi \frac{1}{4} \delta ^{i} _{\rho} \frac{k ^{\prime} _{i} - C _{i}}{ \vert {\gv k} ^{\prime} - {\gv C} \vert ^{3} } . \label{Residues}
\end{equation}
Substitution of the residues (\ref{Residues}) in Eq. (\ref{29}) determines $I ^{(2)} _{\rho} (C) $, and combining the result with  Eq. (\ref{23}) it follows that the vacuum polarization becomes
\begin{align}
{ \Pi} _{A, \, \chi } ^{\mu \nu} (p) &= i \chi \, \frac{e ^{2}}{2} (\det m) (m^{-1}) ^{i}{} _{\lambda} p _{\kappa} \epsilon ^{\mu \nu \lambda \kappa} \notag \\  & \phantom{=} \times \sum _{ \xi = \pm 1 } \xi \int \frac{d ^{3} {\gv k}  }{(2\pi )^3} \frac{k ^{\prime} _{i} - C _{i}}{ \vert {\gv k} ^{\prime} - {\gv C} \vert ^{3} }  H ( k _{0 \xi} ) H (  \Lambda - k _{0 \xi} ),  
\end{align}
which includes  the factor $+i$ arising from the  Eq. (\ref{25}).
It is convenient to make the change of variables
\begin{align}
k ^{\prime \prime} _{i} = k ^{\prime} _{i} - C _{i} , \quad d ^{3} {\gv k} ^{\prime \prime} = d ^{3} {\gv k} ^{\prime} = \det( m ) \, d ^{3} {\gv k}
\end{align}
where we used that $\det (m ^{i}{} _{j}) = \det (m ^{\mu}{}_{\nu} )  =\det (m)$ since $m ^{0}{}_{\nu} = \delta ^{0} _{\nu}$. This yields 
\begin{align}
{ \Pi} ^{\mu \nu} _{A, \, \chi } (p) = i \chi \frac{e ^{2}}{2} (m ^{-1}) ^{i}{}_{\lambda} \, p _{\kappa} \, \epsilon ^{\mu \nu \lambda \kappa}\, \big(  I _{i} ^{+} - I _{i} ^{-} \big)
\end{align}
with
\begin{equation}
I ^{\xi} _{i} = \int \frac{d ^{3} {\gv k} ^{\prime \prime}  }{(2 \pi ) ^{3} } \frac{k ^{\prime \prime} _i}{\vert {\gv k} ^{\prime \prime}  \vert ^{3} } \, H ( k _{0 \xi} ) \, H (  \Lambda  - k _{0 \xi} ). \label{34}
\end{equation}
In the new double-primed variables $ k ^{\prime \prime} $ the constraints $0 < k _{0 \xi} < \Lambda $ read
\begin{align}
 0 < -( k ^{\prime \prime} _{j} + C _{j} ) (m ^{-1}) ^{j}{}_{i} \, m ^{i}{}_{0} + C _{0} + \xi \, \vert {\gv k} ^{\prime \prime}  \vert <  \Lambda . 
\end{align}
From the identifications in Eq. (\ref{Zero_Energy}) these constrains become
\begin{align}
0 < \gv{\mathcal{V}} \cdot {\gv k} ^{\prime \prime}  + \xi \, \vert {\gv k} ^{\prime \prime}  \vert + E_0 < \Lambda  . \label{Condition}
\end{align}
It is convenient to perform the 3-momentum integration in spherical coordinates.  Hence we present the condition (\ref{Condition})  as 
\begin{align}
0<| {\gv k} ''| \, \big(|\gv{\mathcal{V}}| \cos \theta + \xi \big) +E_0 <  \Lambda,  \label{38}
\end{align}
Writing $I _{i} ^{\xi}$  as a linear combination of the vectors ${\cal V} _{i}$ and $C _{i}$, one can show that the projection upon $C _{i}$ is zero, so that
\begin{align}
   I ^{\xi} _{i} = N ^{\xi} \, \mathcal{V} _{i} , 
\end{align}
where  $N ^{\xi}$ depends only on $(| \gv{\mathcal{V}} |, E_0,  \Lambda  )$ as
\begin{align}
N ^{\xi}  = \frac{1}{(2 \pi ) ^{2} |\gv{\mathcal{V}}| } \int _{0} ^{\pi} \!\! d \theta \sin \theta \cos \theta  \int _{0} ^{\infty} \!\! d| {\gv k} ''| \, H ( k _{0 \xi} ) H( \Lambda - k _{0 \xi }) .
\end{align}
Note that for $| \gv{\mathcal{V}} | \geq 1$ there exists an angular direction for which $ | \gv{\mathcal{V}} | \cos \theta +  \xi $ is zero whereby the integral over $| {\gv k} ''|$ diverges and which would imply the need of a nontrivial renormalization procedure. Thus, from now on we take the simpler convergent case $|\gv{\mathcal{V}}|<1$ (relevant for type-I WSMs) which yields  $ | \gv{\mathcal{V}} | \cos \theta +  \xi $ with definite sign: positive (negative) for $\xi=+1 (\xi=-1)$. This in turn fixes the inequalities indicated in Eq. (\ref{38}) as
\begin{align}
 \xi = + 1 & : \quad \frac{ -E_0  }{ | \gv{\mathcal{V}} | \cos \theta + 1 } < |{\gv k} ''| < \frac{ \Lambda-E_0  }{ | \gv{\mathcal{V}} | \cos \theta + 1 } \\ \xi = - 1 & : \quad \frac{  \Lambda-E_0 }{ | \gv{\mathcal{V}} | \cos \theta -1 } < | {\gv k} ''| < \frac{  -E_0 }{ | \gv{\mathcal{V}} | \cos \theta -1 }
\end{align}
The above equations demand us to distinguish the signs of  
$  E_0$ and of $\Lambda - E_0 $, since we must ensure that  $| {\gv k} ''| \geq 0 $. This leaves us with four cases according to the combinations $ {\Lambda -  E_0} \gtrless 0$ and $ E_0 \gtrless 0$. Here we make use of the assumption that $\Lambda>0$, as it constrains the possibilities of these four cases. Under this condition a detailed analysis yields the global result
\begin{align}
 I^{+} _{i} -I ^{-} _{i}= \frac{\Lambda}{\pi ^{2} } N _{\chi} {\mathcal{V}} _{i}  ,   
\quad N_\chi= \frac{1 }{2|\gv{\mathcal{V}}|^3}\Big(|\gv{\mathcal{V}}|-\mathrm{arctanh}
    (|\gv{\mathcal{V}}|)\Big).
\end{align}
The resulting contribution to the vacuum polarization is
\begin{equation}
\Pi ^{\mu \nu} _{A} = i \frac{e ^{2}}{2 \pi ^{2}}  p _{\kappa} \, \epsilon ^{\mu \nu \lambda \kappa} \, \Big[\Lambda_R\, N _{R} \, {\mathcal{V}} _{Ri} \,  (r ^{-1}) ^{i}{}_{\lambda}  - \Lambda_L \, N _{L} \, {\mathcal{V}} _{Li} \, (l ^{-1}) ^{i}{}_{\lambda}    \Big],
\end{equation}
where $\Lambda$ has acquired a chirality dependence through the splitting of the zero excitation energy modes given by
\begin{equation}
    E_{0L(R)} = \gv{\mathcal{V}}_{L(R)}\cdot {\gv C_{L(R)}} + C _{0L(R)}, \quad \Lambda_{L(R)} = \mu-E_{0L(R)}.
    \label{47}
\end{equation}
We thus obtain the additional contribution
\begin{align}
{\cal B}^{(2)}_\lambda=  - \Big[\Lambda_{R} \, N_R \, {\mathcal{V}}_{Ri} \,  (r^{-1})^i{}_\lambda  - \Lambda_L \, N_L \, {\mathcal{V}}_{Li} \, (l^{-1})^i{}_\lambda    \Big] \label{48}
\end{align}
to the coupling $\Theta(x)$ in Eq. (\ref{BLAMBDA}). 

Summarizing, the full effective electromagnetic action of the system described by the fermionic action (\ref{ACTION1}), with the only restrictions $c ^{0}{}_{\nu} =\delta ^{0} _{\nu}$,  \,  $d ^{0}{}_{\nu} = 0 $,  $l ^{\mu}{}_{\nu}$ and $r ^{\mu}{}_{\nu}$ invertible and $| \gv{\mathcal{V}} | < 1 $, is given by the action (\ref{02}) with $\Theta(x) = x ^{\lambda} \big( {\cal B} ^{(1)} _{\lambda} + {\cal B} ^{(2)} _{\lambda} \big)$. Notice that the contribution proportional to $\mu$  is not regularization dependent.  The coupling $\Theta$ is CPT and PT odd
as reflected in Eqs. (\ref{EFS}) with ${\tilde b}_\mu \rightarrow {\cal B}_\mu$ since ${\cal B}_i$ breaks  T but not  P, while ${\cal B}_0$ does the opposite. Then, even if we start  with only the CPT even part of the Lagrangian in Eq. (\ref{ACTION1}) (the $\Gamma^\mu$ term ) by setting $a_\mu=b_\mu=0$ it is not surprising that we obtain a non-zero PT odd $\Theta$  because $\Gamma^\mu$ already includes both  PT even and odd contributions. In this particular case ${\cal B}^{(1)}_\lambda =0, \, \Lambda_\chi=\mu $, but ${\cal B}^{(2)}_\lambda \neq 0$ because ${\gv{\cal V}}_R$ and ${\gv{\cal V}}_L$ remain arbitrary. In other words,   the source of the PT (CPT) odd effective electromagnetic action here  is the $\gamma^5$ PT odd  contribution in $\Gamma^\mu$.  

One particularly interesting and simple system takes place when we take $c^j{}_i=\delta^j{}_i$ and $d^j{}_i=0 $, which corresponds to  the case of arbitrary tilting $\gv{V}_R$ and  $\gv{V}_L$, but with equal isotropic Fermi velocity $v_F=1$ at each node. That is, we take
\begin{align}
C _{R \nu} &= a _{\nu} - b _{\nu} , \qquad \qquad \quad C _{L \nu} = a _{\nu} + b _{\nu} ,   \\  r ^{\mu}{}_{\nu} &= \delta ^{\mu}{} _{\nu} + V ^{i}_{R} \, \delta ^{\mu}{}_{i} \,  \delta ^{0}{}_{\nu}   , \qquad l ^{\mu}{}_{\nu} = \delta^{\mu}{}_{\nu} + V ^{i}_{L}\, \delta ^{\mu}{}_{i} \, \delta ^{0}{}_{\nu},  \\ (r ^{-1})^{\mu}{}_{\nu} &= \delta ^{\mu} _{\nu} - V ^{i} _{R} \delta ^{\mu}{}_{i} \delta ^{0}{}_{\nu}  ,  \qquad (l ^{-1})^{\mu}{}_{\nu} = \delta ^{\mu}_{\nu} - V ^{i} _{L} \delta ^{\mu}{}_{i} \delta ^{0}{}_{\nu} .
\end{align}
Under these conditions ${\cal V} ^{i} = V ^{i}$. 
Putting together the 
contributions for $ {\cal B} _{\lambda} = {\cal B} ^{(1)} _{\lambda} + {\cal B} ^{(2)} _{\lambda} $ we obtain
\begin{align}
{\cal B} _{0} &= b_{0} +  {\gv b} \cdot{\gv U}_{+} + {\gv a} \cdot {\gv U} _{-} - \sum _{\chi = \pm 1} \frac{ \chi  \Lambda _{\chi} }{2 |{\gv V} _{\chi} | } \, \Big( | {\gv V} _{\chi} | - \mathrm{arctanh} ( |{\gv V} _{\chi} | ) \Big), \notag \\ 
{\cal B}^{i} &=  b^{i} - \sum _{ \chi = \pm 1} ( V _{\chi})^{i} \, \frac{ \chi  \Lambda _{\chi}  }{ 2 | {\gv V} _{\chi} | ^{3} } \, \Big( | {\gv V} _{\chi} | - \mathrm{arctanh} ( | {\gv V} _{\chi} | ) \Big),
\label{52}
\end{align}
with $ {\gv U} _{\pm} = \frac{1}{2} ( {\gv V} _{-} \pm {\gv V} _{+} ) $ and $\Lambda_\chi = \mu -E_{0\chi}$ according to Eq. (\ref{47}). We recall that $\chi=1$($\chi=-1$) denote the $R(L)$ contributions. Consistently with the property that the axial anomaly is insensitive to LIV modifications~\cite{LUAGA,FIDEL,SALVIO,Scarpelli},  our results (\ref{52}) show that the Pontryagin density remains unchanged, and that the additional LIV terms in Eq. (\ref{GAMMAMU}) which defines the fermionic action (\ref{ACTION1}), as well as the chemical potential,  modify only the $\Theta$ coupling.

In order to check the consistency of our results with a condensed matter approach, we first establish the conditions under which our general model reduces to that of a WSM as described by the Hamiltonian (\ref{03}). Indeed, the equivalence is achieved by setting ${\gv V}_\chi={\gv v}_\chi, \, {\gv U}_\chi={\gv u}_\chi$,  and
\beq
{\gv a}=0, \quad {\gv b}={\gv {\tilde b}}, \quad 
a_0=-{\gv {\tilde b}}\cdot {\gv u}_-, \quad b_0={\tilde b}_0-{\gv {\tilde b}}\cdot {\gv u}_+ 
\eeq
such that $\Lambda_\chi=\mu+ \chi {\tilde b}_0$.  In particular,  for a type-I Weyl semimetal (i.e.  with $v _{\chi}  < v _{F} = 1$) with the tilting ${\gv v} _{\chi}$ parallel to ${\gv {\tilde b}}$,  the semiclassical Boltzmann approach \cite{Kubo2} leads to the same effective action (\ref{02}) with the Weyl node separation shifted by \cite{ENPREP}
\begin{align}
 {{\tilde b}} \; \to  \; {\tilde b} \, - \sum _{ \chi = \pm 1 } \frac{\chi \Lambda _{\chi} }{ 2 \vert {v} _{\chi} \vert ^{2} } \,  \Big[ v _{\chi} - {\rm arctanh}(v _{\chi} ) \Big],\label{8}
\end{align}
where ${\tilde b}=|\gv{\tilde b}|$ and $\Lambda _{\chi} = \mu  \, +  \chi {\tilde b}_{0} > 0 $ is the chemical potential measured from the nodal point.  Clearly,  the spatial components in Eq. (\ref{52}) successfully simplify to the result of Eq.  (\ref{8}).

\section{Summary and conclusions}

We extend the vacuum polarization method of high energy physics, used in the   calculation of  the  electromagnetic response of fermionic systems, to a large class of fermionic couplings  included  in the Standard Model Extension (SME)  describing Lorentz symmetry violations \cite{Kostelecky0}. Emphasis is made in the CPT odd contribution to the effective action which exhibits remnants of the abelian chiral anomaly due to the appearance of the Pontryagin density $\epsilon ^{\mu \nu \rho \sigma}F _{\mu \nu} F _{\rho \sigma}$. Our approach does not rest in a direct manipulation of chiral transformations which induce the chiral anomaly via the corresponding Jacobian \cite{Fujikawa}. Rather, it  is based in the standard perturbative expansion in powers of the electromagnetic potential of the determinant resulting from the integration of the fermions in the functional integral corresponding to the action (\ref{ACTION1}) plus the specific choices (\ref{GAMMAMU}) . This amounts to the calculation of the vacuum polarization tensor to second order in $A _{\mu}$ starting from the exact modified fermion propagators obtained from the SME. A distinguishing feature  of the method is the inclusion of corrections depending upon the chemical potential $\mu$, which enlarge the range of  possible phenomenological applications. This is possible through a systematic application of thermal field theory methods \cite{KAPUSTA}, which is  exemplified here  in the case of the zero temperature limit. The fermionic systems we consider exhibit the axial anomaly and the resulting electromagnetic actions  are fully described by axion electrodynamics  in the form of Eq. (\ref{02}). All corrections induced by the different parameters in the SME enter only in the coupling $\Theta(x) = 2 {\cal B} _{\lambda} x ^{\lambda}$ to the Pontryagin density,  as shown  in our general  expressions (\ref{29a}) and (\ref{48}). This is consistent with the property that the axial anomaly is insensitive to the Lorentz invariance violating parameters of the SME \cite{LUAGA,FIDEL,SALVIO,Scarpelli}.
Along the text we have insisted in the advantage  of extending these high energy physics methods to condensed matter physics. In fact we envisage interesting applications   in the realm of topological quantum matter, where a rich phenomenology in the electromagnetic response can be explored.
Our general results can be applied to type-I Weyl semimetals (WSM)   with arbitrary tilting and anisotropies. A simpler system  results from the restriction to  a WSM with  arbitrary tilting but equal isotropic Fermi velocity at each node, described by the Hamiltonian (\ref{03}). In this setting, our result (\ref{52}) for the vector contribution ${\cal B}^i$ to the $\Theta$ coupling has been validated by a direct calculation of the conductivity using the Boltzmann semiclassical approach, as shown in Eq. (\ref{8}).
 The calculation and proper regularization of the corrections for the case $| \gv{\mathcal{V}} | \geq 1$ (relevant for type-II WSM) is left unsettled for future research. Pending also is the introduction of a chiral chemical potential which is allowed by the chiral structure of the Hamiltonian describing a WSM with two band-crossings, i.e. each node can be held at different chemical potential. We find appealing also to  pursue  the comparison between  the quantum field theory approach and the  semiclassical Boltzmann formalism. In particular, it would be interesting to elucidate the role of the Berry phase in the first strategy \cite{ENPREP}. 
Finally, the recent calculation of the  one-loop Heisenberg-Euler effective action for zero chemical potential in two of the most studied minimal Lorentz-violating extensions of QED
 \cite{PETROV}, motivates   the  challenge of extending the thermal field theory approach to the calculation of the CPT even contributions to the effective electromagnetic action arising from  the fermionic sector of the SME considered in Eq.  (\ref{GAMMAMU}). 
 
\vspace{-0.4cm}

\section*{Acknowledgements}

L.F.U. and A.M.-R. acknowledge support from the project CONACYT (M\'exico) \# CF-428214. A.M.-R. has been partially supported by DGAPA-UNAM Projects \# IA101320 and \# IA102722.  L.F.U. and A.G.A. were supported in part by Project DGAPA-UNAM  \# 103319.

\end{document}